\documentclass[%
 aip,
 jmp,%
 amsmath,amssymb,
 reprint,%
]{revtex4-2}

\usepackage{graphicx}
\usepackage{dcolumn}
\usepackage{bm}
\usepackage{comment}
\usepackage{xcolor}

\begin{document}

\preprint{AIP/123-QED}

\title{Pressure-Strain Interaction: II. Decomposition in Magnetic Field-Aligned Coordinates}

\author{Paul A. Cassak}
\email{Paul.Cassak@mail.wvu.edu}
\author{M. Hasan Barbhuiya}%
\affiliation{%
Department of Physics and Astronomy, 
West Virginia University, Morgantown, WV 26506, USA
}%
\affiliation{%
Center for KINETIC Plasma Physics, 
West Virginia University, Morgantown, WV 26506, USA
}%
\author{H.~Arthur~Weldon}
\affiliation{%
Department of Physics and Astronomy, 
West Virginia University, Morgantown, WV 26506, USA
}%

\date{\today}

\begin{abstract}
In weakly collisional and collisionless magnetized plasmas, the pressure-strain interaction describes the rate of conversion between bulk flow and thermal energy density.  In this study, we derive an analytical expression for the pressure-strain interaction in a coordinate system with an axis aligned with the local magnetic field.  The result is eight groups of terms corresponding to different physical mechanisms that can contribute to the pressure-strain interaction.  We provide a physical description of each term.  The results are immediately of interest to weakly collisional and collisionless magnetized plasmas and the fundamental processes that happen therein, including magnetic reconnection, magnetized plasma turbulence, and collisionless shocks.  The terms in the field-aligned coordinate decomposition are likely accessible to measurement with satellite observations.
\end{abstract}

\maketitle

\section{Introduction}
\label{sec:intro}

The pressure-strain interaction describes the rate of the direct conversion of energy between bulk flow and thermal (internal) energy density in neutral fluids or plasmas \cite{batchelor67}. It is written as $-({\bf P} \cdot \boldsymbol{\nabla}) \cdot {\bf u}$ (with the minus sign), where ${\bf P}$ is the pressure tensor and ${\bf u}$ is the bulk flow velocity. It was underutilized as a quantity of merit in plasma physics until recently \cite{del_sarto_pressure_2016,Yang17,yang_PRE_2017,del_sarto_pressure_2018}.  Since then, it has been the subject of intense scrutiny, primarily because it can be reliably measured using Magnetospheric Multiscale (MMS) mission \cite{Burch16} satellites.  This has made the observational study of energy conversion into thermal energy in systems out of local thermodynamic equilibrium accessible \cite{Chasapis18,Zhong19,Bandyopadhyay20,bandyopadhyay_energy_2021,Wang21,zhou_measurements_2021}.  The pressure-strain interaction has also been studied in numerical simulations of magnetic reconnection (including magnetotail dipolarization fronts) and magnetized plasma turbulence \cite{del_sarto_pressure_2016,Sitnov18,del_sarto_pressure_2018,Du18,Parashar18,Pezzi19,yang_scale_2019,song_forcebalance_2020,Du20,Fadanelli21,Arro21,Yang_2022_ApJ,Hellinger22}.

This study is the second in a three-part series on the pressure-strain interaction. In Ref.~\cite{Cassak_PiD1_2022} (``Paper I''), it was shown that while the commonly-used decomposition of the pressure-strain interaction into the pressure dilatation and the term known as ${\rm Pi-D}$ separates the compressible and incompressible energy conversion \cite{batchelor67}, it does not separate the effects of converging/diverging flow from flow shear. A different decomposition was derived that does separate these effects. A kinetic description of the terms making up the pressure-strain interaction was provided.

In this study, we present a decomposition of the pressure-strain interaction in a coordinate system with an axis aligned with the local magnetic field.  The motivation is that the magnetic field often organizes the dynamics in magnetized plasmas, and therefore the magnetic field-aligned coordinate system can give a more direct indication of the physics at play (see also Ref.~\cite{Yuen20}).  The only other studies we are aware of that organized pressure-strain interaction relative to the magnetic field was a decomposition of the deviatoric pressure into a ``gyrotropic'' and ``non-gyrotropic'' part in PIC simulations \cite{Du18} and MMS observations \cite{zhou_measurements_2021} and studies on energy conversion in a strongly magnetized plasma ({\it e.g.}, Refs.~\cite{Hazeltine13,Hazeltine18}).  The result of the present study is eight sets of terms that can contribute to the pressure-strain interaction due to compression/expansion and flow shear, with the additional result that they can be caused either by a direct strain of the bulk flow relative to the magnetic field or by a strain of the bulk flow caused by the geometry of the magnetic field. (We emphasize that the contribution to pressure-strain interaction due to the geometry of the magnetic field does not imply that the magnetic field itself is the direct cause of heating \cite{del_sarto_pressure_2016}). We discuss the physical causes of each mechanism, which should be useful for interpreting measurements in simulations and satellite observations. (In what follows, we refrain from referring to a contribution to the pressure-strain interaction as ``heating'' or ``cooling'' because there are a number of effects beyond the pressure-strain interaction that can cause a change to the thermal energy density.)  In Ref.~\cite{Barbhuiya_PiD3_2022} (``Paper III''), we display the terms making up the pressure-strain interaction in both Cartesian and field-aligned coordinates for a particle-in-cell simulation of two-dimensional reconnection.  We use the results to identify the physical causes of the conversion of bulk flow energy to thermal energy during the reconnection process.

The layout of this manuscript is as follows.  A derivation of the pressure-strain interaction in magnetic field-aligned coordinates is provided in Sec.~\ref{sec:theory}. The physical explanation of each term is provided in Sec.~\ref{sec:examples}. Section~\ref{sec:discuss} includes a discussion and conclusions.

\section{Theory}
\label{sec:theory}

\subsection{Pressure-Strain Interaction in Magnetic Field-Aligned Coordinates}
\label{sec:decomposition}

Consider a magnetized plasma with magnetic field ${\bf B}$.  If the magnetic field is straight everywhere, the coordinate system can be chosen with a cardinal direction along the field, and the pressure-strain interaction can be decomposed in Cartesian coordinates as discussed in Paper I.  If the magnetic field is not straight everywhere, we employ a local magnetic field-aligned orthonormal coordinate system, also used in Ref.~\cite{Yuen20}. 

We define the unit vectors of the field-aligned coordinate system as the parallel direction ${\bf \hat{b}}$, the curvature direction $\boldsymbol{{\hat \kappa}},$ and the binormal direction ${\bf \hat{n}}$.  (In differential geometry, these vectors are referred to as the tangent ${\bf \hat{t}}$, normal ${\bf \hat{n}}$, and binormal ${\bf \hat{b}}$ directions, respectively; our notation facilitates the identification of the magnetic field direction.)  The parallel unit vector ${\bf \hat{b}} = {\bf B} / B$ is along the local magnetic field, where $B = |{\bf B}|$ is the magnitude of ${\bf B}$. The magnetic field curvature vector $\boldsymbol{\kappa} = ({\bf \hat{b}} \cdot \boldsymbol{\nabla}) {\bf \hat{b}} = \nabla_\| {\bf \hat{b}}$ is defined in the standard way \cite{yang19}, where $\nabla_\| = {\bf \hat{b}} \cdot \boldsymbol{\nabla}$ is the gradient in the parallel direction. The unit vector $\boldsymbol{{\hat \kappa}}$ in the direction of the curvature is defined as $\boldsymbol{{\hat \kappa}} = \boldsymbol{\kappa} / \kappa $, where $\kappa = |\boldsymbol{\kappa}| = 1 / R$ and $R$ is the local radius of curvature of the magnetic field line.  As is known, ${\bf \hat{b}} \cdot \boldsymbol{{\hat \kappa}} = 0$, which follows because $0 = \nabla_\| ({\bf \hat{b}} \cdot {\bf \hat {b}}) = 2 {\bf \hat{b}} \cdot \nabla_\| {\bf \hat{b}} = 2 {\bf \hat{b}} \cdot \boldsymbol{\kappa}$.  The right-handed coordinate system is completed by defining ${\bf \hat{n}} = {\bf \hat{b}} \times \boldsymbol{\hat{\kappa}}$, which is normal to both the magnetic field and the curvature.

We now calculate the pressure-strain interaction $-({\bf P} \cdot \boldsymbol{\nabla}) \cdot {\bf u}$ in field-aligned coordinates.  We let Greek indices $\alpha, \beta, \ldots$ refer to the $b,\kappa,n$ directions, and we let ${\bf e}_\alpha$ be the unit vector in the $\alpha$ direction.  The quantities in the pressure-strain interaction are written in terms of their elements in field-aligned coordinates as ${\bf P} = P_{\alpha \beta} {\bf e}_\alpha {\bf e}_\beta$, $\boldsymbol{\nabla} = {\bf e}_\alpha \nabla_\alpha$, and ${\bf u} = {\bf e}_\beta u_\beta$. Then, the pressure-strain interaction (using the Einstein summation convention) is
\begin{subequations}
\begin{eqnarray}
    -({\bf P} \cdot \boldsymbol{\nabla}) \cdot {\bf u} & = & -[(P_{\alpha \beta} {\bf e}_\alpha {\bf e}_\beta) \cdot ({\bf e}_\gamma \nabla_\gamma)] \cdot ({\bf e}_\delta u_\delta) \\ & = & -P_{\alpha \beta} (\nabla_\alpha u_\beta) - P_{\alpha \beta} u_\delta [{\bf e}_\beta \cdot (\nabla_\alpha {\bf e}_\delta)] \label{eq:psdecompcoordfree}
\end{eqnarray}
\end{subequations}
since ${\bf e}_\alpha \cdot {\bf e}_\beta = \delta_{\alpha \beta}$, where $\delta_{\alpha \beta}$ is the Kroenecker delta. The first term includes compression/expansion and shear in the standard sense of being related to gradients of the bulk flow with respect to the cardinal directions of the coordinate system, while the second term represents what we call geometrical compression/expansion and geometrical shear because they are caused by gradients of the bulk flow due to the geometry of the magnetic field.  We discuss each in turn in what follows, grouping them into eight sets of terms we call $-PS_j$ (with the minus sign) for $j = 1, \ldots, 8$.

For the first term $-P_{\alpha \beta}(\nabla_\alpha u_\beta)$, $\alpha = \beta$ and $\alpha \neq \beta$ are treated separately.  There are three terms with $\alpha = \beta$, given by
\begin{equation}
    -PS_1 = -P_{\|} (\nabla_\| u_\|), 
\end{equation}
where $P_\| = P_{bb} = {\bf \hat{b}} \cdot {\bf P} \cdot {\bf \hat{b}} = P_{kl} b_k b_l$ is the diagonal pressure element in the parallel direction and $k$ and $l$ are indices in Cartesian coordinates and $u_\| = u_b$, and 
\begin{equation}
   -PS_2 = -P_{\kappa \kappa} (\nabla_\kappa u_\kappa) -P_{nn} (\nabla_n u_n).
\end{equation}
$-PS_1$ describes compression/expansion in the parallel direction, while $-PS_2$ describes compression/expansion in the plane normal to the parallel direction. For $\alpha \neq \beta$, we collect two of the six terms as
\begin{equation}
    -PS_3 = -P_{\kappa b} (\nabla_\kappa u_\|) - P_{nb} (\nabla_n u_\|),
\end{equation}
and the other four are
\begin{equation}
    -PS_4 = -P_{b\kappa} (\nabla_\| u_\kappa) - P_{bn} (\nabla_\| u_n) - P_{\kappa n} (\nabla_\kappa u_n) - P_{n \kappa} (\nabla_n u_\kappa).
\end{equation}
$-PS_3$ describes bulk flow velocity shear of the parallel flow in either perpendicular direction; $-PS_4$ describes bulk flow velocity shear of the flow perpendicular to the field that varies in either the parallel (first two terms) or perpendicular (last two terms) direction.

Finally, we need to simplify the geometric term in Eq.~(\ref{eq:psdecompcoordfree}), which depends on directional derivatives of unit vectors ${\bf e}_\beta \cdot (\nabla_\alpha {\bf e}_\delta)$ and is related to the Christoffel symbol in differential geometry. We first consider parallel gradients of each of the unit vectors.  These are given by the Frenet-Serret formulae from differential geometry \cite{Struik1961,patrikalakis2002shape,Yuen20}, which in our notation are
\begin{subequations}
  \begin{eqnarray}
\nabla_\| {\bf \hat{b}} & = & \kappa \boldsymbol{\hat{\kappa}}, \label{eq:fs1} \\
\nabla_\| \boldsymbol{\hat{\kappa}} & = & \tau {\bf \hat{n}} - \kappa {\bf \hat{b}}, \label{eq:fs2} \\
\nabla_\| {\bf \hat{n}} & = & -\tau \boldsymbol{\hat{\kappa}}, \label{eq:fs3}
  \end{eqnarray}
\end{subequations}
where $\tau = - \boldsymbol{\hat{\kappa}} \cdot \nabla_\| {\bf \hat{n}}$ is the torsion, which is a measure of the degree to which the magnetic field line is not confined to a plane. While $\kappa$ is non-negative by definition, $\tau$ can be positive, negative, or zero.  The first relation follows by definition of $\boldsymbol{\kappa}$. To get the form of the second, one notes $0 = \nabla_\| (\boldsymbol{\hat{\kappa}} \cdot \boldsymbol{\hat{\kappa}}) = 2\boldsymbol{\hat{\kappa}} \cdot \nabla_\| \boldsymbol{\hat{\kappa}}$, so $\nabla_\| \boldsymbol{\hat{\kappa}}$ can only have components in the ${\bf \hat{b}}$ and ${\bf \hat{n}}$ directions.  The third then follows from writing $\nabla_\| {\bf \hat{n}} = \nabla_\| ({\bf \hat{b}} \times \boldsymbol{\hat{\kappa}}) = {\bf \hat{b}} \times \nabla_\| \boldsymbol{\hat{\kappa}}$ and using Eq.~(\ref{eq:fs2}).  Finally, taking $\nabla_\| \boldsymbol{\hat{\kappa}} = \nabla_\| ({\bf \hat{n}} \times {\bf \hat{b}})$ and simplifying gives Eq.~(\ref{eq:fs2}).  A key point is that the local geometry of the magnetic field is determined fully from the curvature $\kappa$ and the torsion $\tau$.  Using the Frenet-Serret formulae provides all the directional derivatives in the parallel direction; the four non-zero ones are 
\begin{subequations}
\begin{eqnarray}
\boldsymbol{\hat{\kappa}} \cdot \nabla_\| {\bf \hat{b}} = -{\bf \hat{b}} \cdot \nabla_\| \boldsymbol{\hat{\kappa}} & = & \kappa, \label{eq:pardirder1} \\ 
{\bf \hat{n}} \cdot \nabla_\| \boldsymbol{\hat{\kappa}} = -\boldsymbol{\hat{\kappa}} \cdot \nabla_\| {\bf \hat{n}} & = &\tau,
\label{eq:pardirder2}
\end{eqnarray}
\end{subequations}
and the other five combinations all vanish.

We also need the directional derivatives in the direction of $\boldsymbol{{\hat{\kappa}}}$ and ${\bf \hat{n}}$.  To find them, define the path length along the magnetic field line as $s$. Then, the coordinates of the magnetic field line can be parametrized by $x(s), y(s),$ and $z(s)$. The derivative of any function $f(s)$ of $s$ in an arbitrary Cartesian direction is given by the chain rule as $\partial f(s) / \partial r_j = (\partial s / \partial r_j) (df/ds)$. This allows us to calculate the directional derivatives of $f(s)$ as
\begin{subequations}
\begin{eqnarray}
\nabla_\| f(s) & = & {\bf \hat{b}} \cdot \boldsymbol{\nabla} f(s) = \frac{df(s)}{ds}, \\
\nabla_\kappa f(s) & = & \boldsymbol{\hat{\kappa}} \cdot \boldsymbol{\nabla} f(s) =
(\boldsymbol{\hat{\kappa}} \cdot \boldsymbol{\nabla} s) \frac{df(s)}{ds}, \\
\nabla_n f(s) & = & {\bf \hat{n}} \cdot \boldsymbol{\nabla} f(s) = ({\bf \hat{n}} \cdot \boldsymbol{\nabla} s )\frac{df(s)}{ds}.
\end{eqnarray}
\end{subequations}
Each is proportional to $df(s)/ds$, and its coefficient is purely geometrical depending on the trajectory of the magnetic field line.  Defining the vector ${\bf W}$ with components $W_b = {\bf \hat{b}} \cdot \boldsymbol{\nabla} s = 1, W_\kappa = \boldsymbol{\hat{\kappa}} \cdot \boldsymbol{\nabla} s$, and $W_n = {\bf \hat{n}} \cdot \boldsymbol{\nabla} s$, we find the gradients in the curvature and binormal direction are
\begin{subequations}
\begin{eqnarray}
\nabla_\kappa & = & W_\kappa \nabla_\|, \\
\nabla_n & = & W_n \nabla_\|.
\end{eqnarray}
\end{subequations}
Using these results and the parallel derivatives given in Eqs.~(\ref{eq:pardirder1}) and (\ref{eq:pardirder2}), all of the remaining directional derivatives $[{\bf e}_\beta \cdot (\nabla_\alpha {\bf e}_\delta)]$ can be calculated in terms of the curvature $\kappa$ and torsion $\tau$.  For example, $\boldsymbol{\hat{\kappa}} \cdot \nabla_\kappa {\bf \hat{b}} = \boldsymbol{\hat{\kappa}} \cdot W_\kappa \nabla_\| {\bf \hat{b}} = W_\kappa \kappa$ and $\boldsymbol{\hat{\kappa}} \cdot \nabla_n {\bf \hat{n}} = \boldsymbol{\hat{\kappa}} \cdot W_n \nabla_\| {\bf \hat{n}} = - W_n \tau$. Continuing in this manner for all the directional derivatives in the geometrical term in Eq.~(\ref{eq:psdecompcoordfree}), we group terms with like factors of the components of ${\bf u}$ and the geometrical factors $\kappa$ or $\tau$ to give
\begin{subequations}
\begin{eqnarray}
-PS_5 & = & u_\kappa \left( P_\| + P_{\kappa b} W_\kappa + P_{n b} W_n \right) \kappa = u_\kappa P_{b \alpha} W_\alpha \kappa, \\
-PS_6 & = & -u_\kappa \left(P_{bn} + P_{\kappa n} W_\kappa + P_{nn} W_n \right) \tau = -u_\kappa P_{n \alpha} W_\alpha \tau, \\
-PS_7 & = & - u_\| \left( P_{b \kappa} + P_{\kappa \kappa} W_\kappa + P_{\kappa n} W_n \right) \kappa = -u_\| P_{\kappa \alpha} W_\alpha \kappa, \\
-PS_8 & = & u_n \left(P_{b \kappa} + P_{\kappa \kappa} W_\kappa + P_{n \kappa} W_n \right) \tau = u_n P_{\kappa \alpha} W_\alpha \tau,
\end{eqnarray}
\end{subequations}
where we consolidated terms using the vector ${\bf W} = {\bf e}_\alpha ({\bf e}_\alpha \cdot \boldsymbol{\nabla})s =  ({\bf \hat{b}} {\bf \hat{b}} + \boldsymbol{\hat{\kappa}} \boldsymbol{\hat{\kappa}} + {\bf \hat{n}} {\bf \hat{n}})\cdot \boldsymbol{\nabla} s$. We note that each of these terms depend on both diagonal and off-diagonal pressure tensor elements.  We show in Sec.~\ref{sec:examples} that the terms proportional to $u_\kappa$ ($-PS_5$ and $-PS_6$) represent geometrical compression/expansion, while the terms proportional to $u_\|$ and $u_n$ ($-PS_7$ and $-PS_8$) represent geometrical shear.  We emphasize that the dependence on bulk velocity without a spatial derivative in these expressions does not imply a velocity gradient is not needed to have a non-zero pressure-strain interaction; rather, these terms contribute to the pressure-strain interaction because of the geometry of the magnetic field.

\subsection{Example of Torsion for a Helical Magnetic Field}
\label{sec:torsionex}

In preparation for explaining the physical manifestation of each term in the pressure-strain interaction, we present a simple example displaying the physical meaning of the torsion $\tau$.  Consider a circular helical magnetic field ${\bf B}_H$ given in Cartesian coordinates by
\begin{equation}
  {\bf B}_H = - B_0 \frac{y}{\sqrt{x^2+y^2}}{\bf \hat{x}} + B_0
  \frac{x}{\sqrt{x^2+y^2}} {\bf \hat{y}} + B_g {\bf \hat{z}},
\end{equation}
where $B_0 \geq 0$ is the magnitude of the in-plane magnetic field (which is uniform in this case) and $B_g$ is the magnitude of the constant and uniform out-of-plane magnetic field. Then, $B = |{\bf B}_H| = (B_0^2 + B_g^2)^{1/2}$, and straight-forward calculations reveal the unit vectors are
\begin{subequations}
  \begin{eqnarray}
{\bf \hat{b}} & = & - b_0
\frac{y {\bf \hat{x}} - x {\bf \hat{y}}}{\sqrt{x^2+y^2}} + b_g {\bf \hat{z}}, \\ 
\boldsymbol{\hat{\kappa}} & = & \frac{-x{\bf \hat{x}}-y{\bf \hat{y}}}{\sqrt{x^2+y^2}}, \\
{\bf \hat{n}} & = & b_g \frac{y{\bf \hat{x}} -x{\bf \hat{y}}}{\sqrt{x^2+y^2}} + b_0 {\bf \hat{z}},
  \end{eqnarray}
\end{subequations}
where $b_0 = B_0 / B$ and $b_g = B_g / B$.  We note for future reference that the direction of the curvature is in the $xy$ plane (radially in, $-{\bf \hat{r}}$) for this magnetic field, but the parallel and binormal directions have both an in-plane (azimuthal $\boldsymbol{\hat{\theta}}$) and out-of-plane (${\bf \hat{z}}$) component.  

A brief derivation reveals that the curvature and torsion for this magnetic field are $\kappa = b_0^2/r_\perp$ and $\tau = b_0 b_g/r_\perp$, where $r_\perp = \sqrt{x^2+y^2}$ is the perpendicular distance from the $z$ axis.  This exemplifies that $\tau$ is positive for a right-handed helix ($B_g > 0$) and negative for a left-handed helix ($B_g < 0$).  If $B_g = 0$, then $\tau = 0$, and the magnetic field lines are confined to planes.  For this particular magnetic field, the torsion is proportional to the current helicity density through ${\bf B} \cdot {\bf J} = (cB^2/4\pi) \tau$ (in cgs units), where ${\bf J} = (c/4\pi) \boldsymbol{\nabla} \times {\bf B}$ is the current density, but this simple relation does not hold for general magnetic field profiles.

\section{Physical Picture of the Pressure-Strain Interaction in Field-Aligned Coordinates}
\label{sec:examples}

\begin{figure}
\includegraphics[width=5.0in]{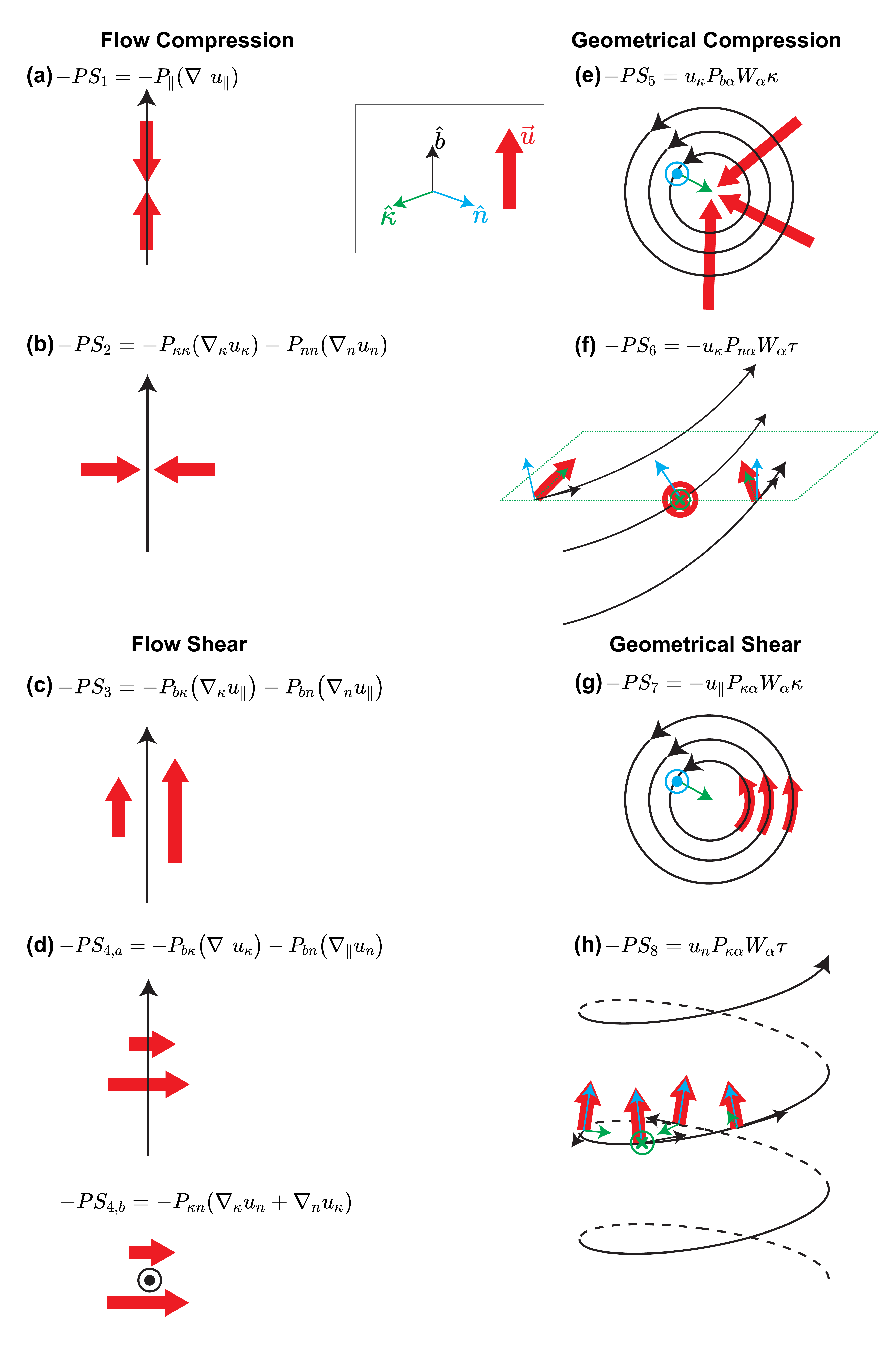}
\caption{\label{fig:2dreconsketches} Representative sketches of the eight sets of terms in the decomposition of the pressure-strain interaction in field-aligned coordinates.  Black arrows represent the magnetic field ${\bf B}$. Green and blue arrows denote the curvature and binormal directions, respectively.  Red arrows denote the bulk flow ${\bf u}$.  The eight sketches represent (a) parallel flow compression $-PS_1$, (b) perpendicular flow compression $-PS_2$, (c) shear of parallel flow in the perpendicular direction $-PS_3$, (d) shear of perpendicular flow in the perpendicular and/or parallel directions $-PS_4$, (e) perpendicular geometrical compression $-PS_5$,  (f) torsional geometrical compression $-PS_6$, (g) parallel geometrical shear $-PS_7$, and (h) torsional geometrical shear $-PS_8$.}
\end{figure}

We now turn to the physical description of the contributions to the pressure-strain interaction in field-aligned coordinates.  The analysis from the previous section showed that there are eight groups of terms. Sketches of representative examples of the general terms are given for each in Fig.~\ref{fig:2dreconsketches}. In each case, the magnetic field ${\bf B}$ is sketched using black arrows, while the bulk flow ${\bf u}$ is represented by red arrows.  The curvature and binormal directions are depicted in green and blue, respectively.  We treat each term in turn.  The fluid description of the Cartesian analogue of the terms that arise in the absence of a heat flux was treated in Refs.~\cite{del_sarto_pressure_2016,del_sarto_pressure_2018}. We emphasize the kinetic description of each term, which complements the fluid description and makes no assumptions about the presence or absence of a heat flux. In so doing, when describing compression/expansion, we phrase it in terms of compression and analogous arguments can be used to describe expansion. We stress that the sketches are intended to give the simplest examples of the terms to illustrate the fundamental mechanism, while not intending to represent the general case. 

\subsection{Parallel Flow Compression/Expansion: $-PS_1 = -P_\| (\nabla_\| u_\|)$}

For $-PS_1 = -P_\| (\nabla_\| u_\|)$, the term describes compression due to a converging flow in the magnetic field direction ${\bf \hat{b}}$.  A converging flow ($\nabla_\| u_\| < 0$) is associated with a positive contribution to the pressure-strain interaction, while a diverging flow ($\nabla_\| u_\| > 0$) is associated with a negative contribution.  It is largely as expected from a fluid treatment, but the departure is that the term only depends on the parallel pressure $P_\|$.  This is sketched in Fig.~\ref{fig:2dreconsketches}(a), showing it for a straight magnetic field line and oppositely directed converging flows.  However, this mechanism operates even for curved magnetic field lines provided there are converging/diverging flows in the parallel direction. Also, converging/diverging flows can occur without a change of direction of the flow.

Kinetically, this term describes a fluid or plasma with an arbitrary phase space density, but only the parallel diagonal pressure element of its pressure contributes.  The mechanism is analogous to parallel compressional heating in a Cartesian coordinate system, as displayed in Fig.~1(a) of Paper I.  Briefly, the phase space density at a point where there is converging flow elongates in the parallel direction, which is the kinetic manifestation of heating, due to the offset of the nearby distributions from the bulk flow at the point of interest.

\subsection{Perpendicular Flow Compression/Expansion: $-PS_2 = -P_{\kappa \kappa} (\nabla_\kappa u_\kappa) - P_{nn} (\nabla_n u_n)$}

For $-PS_2 = -P_{\kappa \kappa} (\nabla_\kappa u_\kappa) - P_{nn} (\nabla_n u_n)$, the two terms describe compression of the bulk flow in the plane perpendicular to the magnetic field.  This again is as expected from a fluid treatment, but again shows that only the diagonal component of the pressure tensor parallel to the converging flow contributes.  This is sketched in Fig.~\ref{fig:2dreconsketches}(b), showing it for a straight magnetic field line.  This effect also occurs regardless of the shape of the magnetic field line provided there are converging flows across it.  The flows need not go to zero at the field line; they only need to converge or diverge. This mechanism is analogous to perpendicular compressional heating in a Cartesian coordinate system, for which the kinetic interpretation was displayed in Fig.~1(b) of Paper I.  As with parallel compression ($-PS_1$), diagonal elements of ${\bf P}$ are non-negative, so compression necessarily has a positive contribution to the pressure-strain interaction and expansion necessarily has a negative contribution.

\subsection{Parallel Flow Sheared in the Perpendicular Direction: $-PS_3 = - P_{b \kappa} \left( \nabla_\kappa u_\| \right) - P_{bn} \left(\nabla_n u_\| \right)$}
\label{sec:pidincompressiblephysics}

For $-PS_3 = - P_{b \kappa} \left( \nabla_\kappa u_\| \right) - P_{bn} \left(\nabla_n u_\| \right)$, these terms describe a parallel flow that varies in the plane normal to the magnetic field. A sketch exemplifying this term is given in Fig.~\ref{fig:2dreconsketches}(c).  This mechanism does not require a curved magnetic field and depends on flow gradients, so it describes a bulk flow shear similar to expectations from a fluid picture.  The key kinetic aspect is that this mechanism is non-zero only if there is a non-zero off-diagonal pressure tensor element in the direction of the magnetic field.  Kinetically, this effect is analogous to flow shear in a Cartesian coordinate system, as displayed in Fig.~2 of Paper I, {\it i.e.,} it is associated with collisionless viscosity, so it can be positive or negative and is in principle reversible. This figure explains why the necessary off-diagonal pressure tensor elements need a component along the magnetic field.  In a weakly collisional or collisionless plasma, the off-diagonal pressure tensor elements can be positive or negative, which determines whether a given flow profile has a positive or negative contribution to the pressure-strain interaction.

\subsection{Perpendicular Flow Shear: $-PS_4 = - P_{b \kappa} \left( \nabla_\| u_\kappa \right) - P_{bn} \left(\nabla_\| u_n \right) - P_{\kappa n} (\nabla_\kappa u_n + \nabla_n u_\kappa)$}
\label{sec:pidincompressiblephysics2}

For $-PS_4 = - P_{b \kappa} \left( \nabla_\| u_\kappa \right) - P_{bn} \left(\nabla_\| u_n \right) - P_{\kappa n} (\nabla_\kappa u_n + \nabla_n u_\kappa)$, the mechanisms are of two related varieties.  They are similar to the heating mechanism in the previous subsection in that they do not rely on any curvature of the magnetic field, and as with $-PS_3$ they require a shear in the bulk flow velocity.  The first two terms describe a bulk flow in the plane perpendicular to the magnetic field that varies in the parallel direction.  A representative sketch is given in the top of Fig.~\ref{fig:2dreconsketches}(d). The second two terms describe a flow perpendicular to the magnetic field that varies in the orthogonal perpendicular direction, as sketched in Fig.~\ref{fig:2dreconsketches}(d) on the bottom.

As with $-PS_3$, these terms correspond to the kinetic notion of collisionless viscosity, requiring off-diagonal pressure tensor elements.  In each case, the off-diagonal pressure tensor element must have a component in the direction of the gradient of the flow. As with $-PS_3$, whether a given term leads to a positive or negative contribution to the pressure-strain interaction depends on the flow shear and the sign of the off-diagonal pressure tensor element in question. Kinetically, the heating mechanism is analogous to the Cartesian coordinate system result displayed in Fig.~2 of Paper I, which explains why the particular off-diagonal pressure-tensor elements in the expressions are needed for this term to be non-zero.  

\subsection{Perpendicular Geometrical Compression/Expansion: $-PS_5 = u_\kappa P_{b\alpha}W_\alpha \kappa$} 
\label{sec:ps6}

For $-PS_5 = u_\kappa P_{b\alpha}W_\alpha \kappa$, the contribution to the pressure-strain interaction requires a curved magnetic field (planar or not) and takes place in the plane of the curvature and the magnetic field.  The bulk flow is in the direction of the curvature, which is perpendicular to the magnetic field lines.  A simple example is shown in Fig.~\ref{fig:2dreconsketches}(f), with a positive $u_\kappa$ that need not vary as one traverses along the magnetic field.  This mechanism is a form of geometrical compression, which results from the red arrows denoting the flow converging in the direction of the curvature, which is the cause of compression in the fluid sense.  Unlike the four terms discussed in the previous subsections, this mechanism does not require a gradient in the bulk velocity component in field-aligned coordinates; instead the flow shear arises due to the curvature of the magnetic field.  

There is an important aspect of this mechanism: it requires at least one of the pressure tensor elements in the ${\bf \hat{b}}$ direction to be non-zero since it is proportional to $P_{b \alpha}$. The reason for this is that if the plasma were perfectly cold in the ${\bf \hat{b}}$ direction, then all motion would be perpendicular to the magnetic field, so it would be in the direction of the flow.  This would not cause any mixing in the direction normal to the flow, so there is no contribution to the pressure-strain interaction for such flow.  If there is random particle motion in the ${\bf \hat{b}}$ direction, mixing can occur that spreads the phase space density in the ${\bf \hat{b}}$ direction, which is therefore associated with a non-zero contribution to the pressure-strain interaction. More specifically, particles with positive or negative $v_\|$ and positive $v_\kappa$ on the outer magnetic field line move inward to the middle field line in time, providing a population with non-zero $v_\| > 0$ and $v_\kappa > 0$.  Particles on the inner magnetic field line with non-zero $v_\|$ and negative $v_\kappa$ move outward to the middle field line, providing a population with non-zero $v_\|$ and $v_\kappa <0$.  This changes the spread in the distribution at the middle field line, which is the kinetic manifestation of a change to the thermal energy. This mechanism for geometrical compression can lead to a positive or negative contribution to the pressure-strain interaction depending on the phase space density of the plasma, in contrast to bulk flow compression $-PS_1$ or $-PS_2$ which necessarily makes a positive contribution to the pressure-strain interaction.

\subsection{Torsional Geometrical Compression: $-PS_6 = -u_\kappa P_{n\alpha} W_\alpha \tau$}

For $-PS_6 = -u_\kappa P_{n\alpha} W_\alpha \tau$, the mechanism is related to perpendicular geometrical compression discussed in the previous subsection, but it requires a non-planar magnetic field, {\it i.e.,} a magnetic field with a non-zero torsion.  The key is that in the absence of torsion, having a bulk flow in the curvature direction $u_\kappa$ means that the flow converges in the plane of the curvature and magnetic field, as shown in Fig.~\ref{fig:2dreconsketches}(e) for $-PS_5$.  Thus, all the plasma that is converging comes from the same plane initially.

As an example of this mechanism in a magnetic field that has non-zero torsion, consider the simple case of a helical right-handed torsional magnetic field of the type discussed in Sec.~\ref{sec:torsionex}. The oblique view in Fig.~\ref{fig:2dreconsketches}(f) shows the magnetic field twisting out of the plane, with the dashed line representing the $xy$ plane that contains the curvature vector $\boldsymbol{\kappa}$.  If the flow in the $\boldsymbol{\kappa}$ direction is converging, the particles in the $xy$ plane that end up in the region of converging flow originate from regions of the magnetic field that are separated in the binormal ${\bf \hat{n}}$ direction.  If the plasma is perfectly cold in the ${\bf \hat{n}}$ direction, the phase space density in the region of converging flow does not broaden and thus there is no contribution to the pressure-strain interaction due to shear [although there can be a contribution due to the ${\bf \hat{b}}$ direction from $-PS_5$, as sketched in Fig.~\ref{fig:2dreconsketches}(e)]. If, however, there are any particles with random velocity in the ${\bf \hat{n}}$ direction, there is mixing in that direction and a compressional effect gives a non-zero contribution to the pressure-strain interaction.  This mechanism is purely due to the geometry of the magnetic field, so we call it torsional geometrical compression. This term is not positive definite as with the other geometrical terms, so it can be associated with positive or negative contributions to the pressure-strain interaction.

\subsection{Parallel Geometrical Shear: $-PS_7 = -u_\| P_{\kappa \alpha} W_\alpha \kappa$} 

For $-PS_7 = -u_\| P_{\kappa \alpha} W_\alpha \kappa$, the mechanism requires a magnetic field line with curvature (planar or not) when there is a bulk flow with a component parallel or anti-parallel to the magnetic field.  An example of this for a circular magnetic field line with $u_\| > 0$ is sketched in Fig.~\ref{fig:2dreconsketches}(g). In the fluid sense, this mechanism leads to shear because the inner field lines are shorter than the outer field lines, so a flow profile with uniform $u_\|$ implies that there is a shear due to the plasma traversing the shorter curved field line further in azimuthal angle than along the longer field lines.

To understand this mechanism in kinetic theory, first suppose none of the particles have any velocity component in the $\boldsymbol{\hat{\kappa}}$ direction, {\it i.e.}, the plasma is perfectly cold in the $\boldsymbol{{\hat {\kappa}}}$ direction. This implies that all the motion of the particles is confined to the magnetic surfaces.  In this case, in the next small increment in time, the particles remain confined to the magnetic surfaces.  Therefore, there is no mixing of particles between different magnetic surfaces, and there is no contribution to pressure-strain interaction.  This is why $-PS_7$ is proportional to $P_{\kappa \alpha}$, {\it i.e.,} why random motion in the $\boldsymbol{\hat{\kappa}}$ direction is necessary for this mechanism to occur.

Now consider a phase space density such that there is a non-zero $P_{\kappa \kappa}$, which means that particles have some random motion in the direction perpendicular to the magnetic surfaces.  Consider the time evolution of a phase space density at the middle of the three magnetic field lines in Fig.~\ref{fig:2dreconsketches}(g).  As the particles go around the curve in the magnetic field lines, particles with positive  velocity $v_\kappa$ on the inner field line in Fig.~\ref{fig:2dreconsketches}(g) move outward in the next increment in time, showing up as a positive $v_\kappa$ population at the middle field line.  Similarly, particles with negative $v_\kappa$ on the outer field line move inward, showing up as a negative $v_\kappa$ population at the middle field line.  This broadens the phase space density at the middle field line in the $\boldsymbol{{\hat \kappa}}$ direction, which is associated with an increase in thermal energy in the kinetic sense. This is an effective shear due to the geometry of the magnetic field, so we refer to it as parallel geometric shear.  We note as an application that this mechanism is important for a plasma in a magnetic mirror configuration.

There are analogous mechanisms for distributions with non-zero $P_{b \kappa}$ and $P_{\kappa n}$.  Since these off-diagonal pressure tensor elements can be positive or negative, shear due to field line geometry can contribute to a positive or negative contribution to the pressure-strain interaction. 

\subsection{Torsional Geometrical Shear: $-PS_8 = u_n P_{\kappa \alpha} W_\alpha \tau$}

For $-PS_8 = u_n P_{\kappa \alpha} W_\alpha \tau$, the mechanism relies on the magnetic field having torsion, but the flow is in the binormal direction. We again appeal to the simple example of the right-handed circular helical field discussed in Sec.~\ref{sec:torsionex}.  To isolate the effect of the torsion, we consider a flow with uniform $u_n$.  (If it were not uniform, there would be a perpendicular flow shear as in $-PS_4$ in addition to the geometrical shear.)  A sketch of this is in Fig.~\ref{fig:2dreconsketches}(h). It was pointed out in Sec.~\ref{sec:torsionex} that the binormal direction ${\bf \hat{n}}$ has components both in the axial ${\bf \hat{z}}$ and azimuthal $\boldsymbol{\hat{\theta}}$ directions. This can be seen from the red arrows in the sketch.  The component of the bulk flow in the ${\bf \hat{z}}$ direction does not introduce shear because $u_n$ is the same on all magnetic field lines in this configuration.  Thus, the axial part is not associated with a contribution to the pressure-strain interaction. 

However, the component in the $\boldsymbol{\hat{\theta}}$ direction describes flow in the azimuthal direction.  As with $-PS_7$, the geometry of the curved magnetic field lines imposes that there is a shear effect on the plasma because fluid elements with the same $u_\theta$ on different magnetic surfaces traverse the circular cross-section of the magnetic surface more rapidly for smaller magnetic surfaces than larger magnetic surfaces. If the phase space densities are perfectly cold in the $\boldsymbol{\hat{\kappa}}$ direction, then all particles are confined to the magnetic surfaces, and therefore there is no mixing and no contribution to the pressure-strain interaction. If, however, there are particles with a random $v_\kappa$, particles on different magnetic surfaces mix according to a flow shear-like mechanism in the kinetic description,  so there is a change to the thermal energy density.  This is why $-PS_8$ is proportional to $P_{\kappa \alpha}$. We refer to this as torsional geometrical shear.  It can be positive or negative depending on the torsion, pressure tensor elements, and the flow direction.

\section{Discussion and Conclusions}
\label{sec:discuss}

In this study, we derive an expression for the pressure-strain interaction, the term describing the rate of conversion of energy between bulk flow and thermal energy density, in magnetic field-aligned coordinates for use in magnetized plasmas. As expected, there are contributions related to compression/expansion and bulk flow shear.  However, in field-aligned coordinates, each effect has contributions directly from the spatial dependence of the bulk velocity itself as well as contributions from velocity shear caused by the geometry imposed by the path of the magnetic field line.  It is important to stress that the magnetic field itself and magnetic forces do not cause the pressure-strain interaction to be non-zero \cite{del_sarto_pressure_2016}, it is simply the flow pattern relative to the magnetic fields that contribute.  The geometric compression/expansion and geometric shear are parametrized in terms of the magnetic field curvature $\kappa$ and torsion $\tau$.  The former is well-known in plasma physics; the latter is borrowed from differential geometry and is much less employed in plasma physics \cite{Yuen20}, and describes the extent to which the local magnetic field deviates from lying in a plane.

We provide a picture of the physical effects contributing to the pressure-strain interaction using the kinetic theory description for each of the sets of terms $-PS_1$ through $-PS_8$ that arise from the analysis. The fluid description for plasmas in the absence of a heat flux was previously provided \cite{del_sarto_pressure_2016,del_sarto_pressure_2018}.  We emphasize that the two descriptions complement each other and must agree with each other when the same approximations are made in both pictures. The kinetic approach discussed here makes no assumptions about the presence of a heat flux.  The physical mechanism of the parallel and perpendicular compression/expansion $-PS_1$ and $-PS_2$, and the shear in the bulk flow $-PS_3, -PS_{4,a}$, and $-PS_{4,b}$ are analogous to compression/expansion and bulk velocity shear in Cartesian coordinates \cite{del_sarto_pressure_2018,Cassak_PiD1_2022}.  In the kinetic description, the physical mechanism for geometric compression and geometric shear is random motion in the direction perpendicular to the flow in $-PS_5$ through $-PS_8$, which causes mixing of particles that gives rise to a pressure-strain interaction contribution.

We expect these results, especially the simple sketches of the physical contributions to pressure-strain interaction in Fig.~\ref{fig:2dreconsketches}, will be useful in studying energy conversion in weakly collisional and collisionless magnetized plasmas.  Physical systems where the pressure-strain interaction has been used to study energy conversion, and where the present results may be useful, include plasma turbulence and magnetic reconnection.  We expect it to also be useful for the study of collisionless shocks.  The quantities derived here can be readily calculated in kinetic simulations (particle-in-cell and Vlasov/Boltzmann in particular) of these phenomena.  Moreover, recent observational studies \cite{Cozzani19,Qi19,Bandyopadhyay20b,Huang20,Rogers21} using the MMS satellites have directly measured the magnetic field curvature $\kappa$, so this quantity of importance for the present study is accessible to measurement.  We are unaware of any calculations of the magnetic field line torsion $\tau$ using satellite data, but it is a simple extension of calculating the magnetic field and curvature directions, so it should be able to be calculated.

We make three important points about the present results.  First, the pressure-strain interaction $-({\bf P} \cdot \boldsymbol{\nabla}) \cdot {\bf u}$ is a scalar quantity, meaning it is invariant in different coordinate systems.  Thus, whether the pressure-strain interaction is calculated in Cartesian coordinates (Paper I) or field-aligned coordinates, it remains the same. However, there is mixing between compression/expansion and flow shear when changing coordinate systems. Second, the results here rigorously provide the pressure-strain interaction contributions in field-aligned coordinates for applications to magnetized plasmas.  However, we stress there are settings in magnetized plasmas where particles become demagnetized, and the direction of the magnetic field no longer organizes the dynamics \cite{Servidio12,Burch16b,Egedal18}.  Thus, caution is necessary to not assume the magnetic field direction is necessarily the direction that best organizes a general pressure tensor.  Finally, the decomposition in field-aligned coordinates presented here does not use any properties of the magnetic field itself.  Thus, if an application arises for which there is a different preferred direction other than the magnetic field, the analysis presented here remains valid with ${\bf \hat{b}}$ simply becoming the preferred direction.  This underscores the key point that the magnetic field and magnetic forces themselves do not give rise to the pressure-strain interaction, it is only the bulk flow gradients relative to the geometry set up by the magnetic field that gives rise to the pressure-strain interaction.

In Paper III, we use the results obtained here and in Paper I to analyze the mechanisms by which the pressure-strain interaction describes the conversion of bulk flow energy density to thermal energy density during magnetic reconnection using two-dimensional particle-in-cell simulations.  For future work, it would be interesting to employ the decomposition of the pressure-strain interaction discussed here in observational data, especially using the MMS satellites.  Applications of the results to plasma turbulence and collisionless shocks, as well as other manifestations of reconnection including three-dimensional systems, would also be very interesting.

\begin{acknowledgments}
We acknowledge beneficial conversations with Yan Yang.  We gratefully acknowledge support from NSF Grant PHY-1804428, DOE grant DE-SC0020294, and NASA grant 80NSSC19M0146.  This research uses resources of the National Energy Research Scientific Computing Center (NERSC), a DOE Office of Science User Facility supported by the Office of Science of the US Department of Energy under Contract no.~DE-AC02-05CH11231.
\end{acknowledgments}


\providecommand{\noopsort}[1]{}\providecommand{\singleletter}[1]{#1}%

\end{document}